\documentclass[12pt,preprint]{aastex}

\def\mbf#1{\mbox{\boldmath ${#1}$}}

\slugcomment{Accepted for publication in The Astrophysical Journal}

\begin{document}


\title{Thermal Response of A Solar-like Atmosphere to \\
    An Electron Beam from A Hot Jupiter: A Numerical Experiment}


\author{Pin-Gao Gu}
\affil{Institute of Astronomy \& Astrophysics, Academia Sinica,
    Taipei 10617, Taiwan} \email{gu@asiaa.sinica.edu.tw}

\author{Takeru K. Suzuki}
\affil{School of Arts \& Sciences, University of Tokyo, Komaba,
Meguro, Tokyo, 153-8902, Japan} \email{stakeru@ea.c.u-tokyo.ac.jp}




\begin{abstract}
We investigate the thermal response of the atmosphere of a
solar-type star to an electron beam injected from a hot Jupiter by
performing a 1-dimensional magnetohydrodynamic numerical
experiment with non-linear wave dissipation, radiative cooling,
and thermal conduction. In our experiment, the stellar atmosphere
is non-rotating and is modelled as a 1-D open flux tube expanding
super-radially from the stellar photosphere
to the planet. An electron beam is assumed to be generated from
the reconnection site of the planet's magnetosphere. The effects
of the electron beam are then implemented in our simulation as
dissipation of the beam momentum and energy at the base of the
corona where the Coulomb collisions become effective. When the
sufficient energy is supplied by the electron beam, a warm region
forms in the chromosphere.
This warm region greatly enhances the radiative fluxes
corresponding to the temperature of the chromosphere and
transition region. The warm region can also intermittently
contribute to the radiative flux associated with the coronal
temperature due to the thermal instability. However, owing to the
small area of the heating spot, the total luminosity of the
beam-induced chromospheric radiation is several orders of
magnitude smaller than the observed Ca II emissions from HD
179949.
\end{abstract}





\section{Introduction}

Hot Jupiters are Jupiter-mass planets located within $\sim$ 0.1 AU or less from
their parent stars. Because of the close proximity to the parent stars, hot
Jupiters have been expected to be able to influence their stellar companions via
magnetic \citep{Cuntz,RS00} and/or tidal interactions
\citep{Lin96,Jackson09,Pfahl08}. The observations of Ca II H \& K lines from a
number of stars harboring a hot Jupiter have suggested that the chromospheric
activities, characterized either by line intensity or short-time variability,
sometimes correlate with the orbits of their planets \citep{Shk03,Shk05,Shk08}.
These phenomena can be modelled as a hot spot or a more ``variable" region, despite
residing in the chromosphere, following the planet's orbital motion with a phase
difference. In particular, the observations carried out in 2001, 2002, and 2005
imply that a hot spot on HD 179949 persistently leads the planet by $\sim 70^\circ$
with the intensity of $\sim 10^{27}$ erg/s in Ca II emissions. The similar phase
lead of a variable region in optical has been suggested by the MOST satellite
photometry for the hot-Jupiter host star $\tau$ Bootis \citep{Walker08}. Since the
planet-induced stellar activities occur only once during one orbital period, the
origins of this ``spot" have been attributed to magnetic rather than tidal
interactions.

One of the commonly adopted scenarios to describe the star-planet
magnetic interactions is the magnetic interactions between Jupiter
and its Galilean satellites (see Zarka 2007 for a review). In this
scenario, the orbital motion of the Galilean satellites relative
to Jupiter's magnetosphere taps the orbital energy of the
satellites at a rate that depends on detailed modelling on the
magnetic interactions. The interactions can be classified into two
types: the unipolar interaction with an unmagnetized satellite
such as Io and the magnetic reconnection with a magnetized
satellite such as Ganymede. The energy is then transported by the
Alfv\'en waves and/or by a fast electron beam along the field
lines from the satellites to Jupiter's surface where the energy is
dissipated, thereby explaining the satellite-induced emissions
from Jupiter. In the case of star-planet interactions, this
picture has been modified to take into consideration stellar winds
along open field lines or to allow for a large static magnetic
loop connecting the star and the planet. As a result, the phase
differences between the stellar ``spot" and the planet are
explained by the time lag due to the Alfv\'en travel time to the
star in the Alf\'ven-wave model \citep{Pre06}, or by the large
magnetic loop having a geometry across longitudes of the star in
the electron-beam case \citep{Lanza08}. A three-dimensional
resistive magnetohydrodynamic simulation was performed to study
how the magnetic field-aligned current can develop from a hot
Jupiter \citep{Pre07}.

However, how the stellar atmosphere thermally responds to any
energy injection from the planet so as to generate the
chromospheric emissions of $\gtrsim 10^{27}$ erg/s remains
elusive. The magnetic energy at the magnetopause of a hot Jupiter
has been estimated to be insufficient to supply the energy rate of
$\gtrsim 10^{27}$ erg/s \citep{Shk05,zak07} if the strength of the
stellar surface field is $\sim$ a few Gauss \citep{Catala}.
Furthermore, if this energy dissipation rate arises entirely from
the orbital energy of a hot Jupiter, most hot Jupiters would
plunge into their central stars in a few billion years, a
timescale comparable to the age of these planetary systems
\citep{Shk05}. However, the gas in a stellar atmosphere is
certainly not quiescent but is fluctuating with the free energy
that may be liberated as a hot Jupiter encounters the turbulent
stellar fields along its orbit. \citet{Cuntz} and \citet{Saar}
took into account the energy contribution from the stellar
macroturbulence velocity and compared the strength of the
planet-star interactions relative to each other for a number of
hot-Jupiter systems. \citet{Gu05} postulated that most of the
planet induced emissions may result from stellar turbulent energy
to reconcile the energetic problem. To further examine this
possibility, a more realistic stellar atmosphere model implemented
with an energy equation involving stellar turbulence is required.

In the case of the solar atmosphere, the coronal fields that can
reach the typical orbits of hot Jupiters are the open magnetic
fields emanating from the coronal holes. The dynamical features of
the open fields are the non-linear fluctuations in the corona and
the high-speed winds (800 km/s at $\approx 1$ AU). To explain
these features, \cite{SI05} (hereafter SI05) introduced a
1-dimensional magnetohydrodynamic simulation with radiative
cooling and thermal conduction. In this simulation, the authors
have self-consistently treated the transfer of
mass/momentum/energy by solving the magnetic waves propagating
from the photosphere to the interplanetary space. The open field
lines are assumed to expand super-radially from the photosphere to
the corona \citep{KH}, which is consistent with the
spectropolarimetric measurements of magnetic structures in the
Sun's polar region \citep{Tsuneta}. The heating and acceleration
of the gas in the coronal holes are achieved by the dissipation
and the pressure of the non-linear magnetic waves. The model can
explain the structure of the solar atmosphere (i.e. consisting of
the chromosphere, the thin transition region, and the corona) as
well as the high-speed winds from the coronal holes.

In this paper, we employ the magnetohydrodynamic simulation described above and
model the open fields as a 1-dimensional flux tube from the stellar photosphere to
the hot Jupiter. The aim is to conduct a numerical experiment as a starting point
to study the thermal properties of the stellar chromosphere, transition region, and
corona in response to an energy injection. Our study concerns HD 179949, but it is
also guided by information about the solar wind due to the fact that various
properties of HD 179949, especially in regard to its chromosphere, corona and wind,
are poorly constrained. In \S2, we describe the stellar atmosphere model in the
presence of an inward-propagating electron beam. The numerical results are
presented in \S3. The paper concludes with a summary and discussion in \S4.


\section{Model Description}

We consider a magnetized hot Jupiter revolving in a circular orbit
on the equatorial plane of its parent star.
As the planet orbits through the open flux tubes emanating from
the stellar surface, magnetic reconnections occur and generate
plasma jets propagating inwards along the flux tubes to the parent
star. In the following subsections, we shall describe the model of
electron-beam injection and the stellar atmosphere associated with
the flux tube.

\subsection{Electron-Beam Injection}
\label{sec:ebi}

Since HD 179949 has been the canonical planetary system
highlighted by most previous studies for magnetic interactions, we
adopt the parameters of HD 179949 as an illustrative example for
our main study. In other words, we focus on a hot Jupiter orbiting
at the radial distance $r=7.8R_*$ around a central star of mass
$M_*=1.21M_{\odot}$, radius $R_*=1.22R_{\odot}$, and effective
temperature $T_{\rm eff}=6168$ K \citep{Butler}. In terms of
$M_*$, $R_*$, and $T_{\rm eff}$,
these stellar parameters are noticeably different from the solar values, although
in the broad sense HD 179949 still constitutes a solar-type star.

The typical field strength at the plane orbit is $B\sim 0.1-0.01$
G, for the {\it average} radial field strength $\sim 1-10$ G at
the stellar surface (i.e. $B\propto r^{-2}$). At $r\approx
10R_{\odot}$, the planet is located inside the Alfv\'en radius of
the star (see the next subsection) and therefore no fast MHD
shocks form as the stellar winds encounter the planet's
magnetosphere (cf. Zarka 2001; Ip et al. 2004; Preussue et al.
2005). Assuming a dipole field for the planet's magnetic field, we
can estimate the radius of the magnetopause $R_{mp}$ at which the
stellar and the planet's fields are balanced:
\begin{equation}
R_{mp}=4.64 R_J \left({R_p\over R_J}\right) \left[ \left({B_p
\over 1\,{\rm G}} \right) \left( {B_* \over 1\,{\rm G}}
\right)^{-1} \left( {a \over 10R_{\odot}} \right)^2 \left( {R_*
\over R_{\odot}} \right)^{-2} \right]^{1/3}, \label{eq:R_mp}
\end{equation}
where $B_p$ and $B_*$ are the average surface fields of the planet
and the parent star respectively, $R_J$ is the radius of Jupiter,
and $a$ is the semi-major axis. As can be clearly seen from the
above equation, $R_{mp}<<a=10R_{\odot}$. Therefore the magnetic
field strength $B_{mp}$ at the magnetopause is almost equal to the
stellar field strength at $r=a$; i.e., $B_{mp}\sim 0.1-0.01 G$.

We consider a jet originating from the reconnection sites at the
magnetopause. The energy flux liberated by the reconnection is on
the order of $~ (B_{mp}^2/4\pi)v_{rel}$, where $v_{rel}$ is the
speed of the planet relative to the stellar fields. For a slowly
rotating solar-type star, $v_{rel}$ is close to the orbital speed
of the planet $\sqrt{GM_\odot /a}=1.38\times 10^7$ cm/s (cf.
$v_{rel}\gtrsim 300$ km s$^{-1}$ in \citet{zak07}). Thus, we
arrive at a rough estimate of the liberated energy flux
\begin{equation}
\frac{B_{mp}^2}{4\pi}v_{rel} \approx 10^2 {\rm erg\;
cm^{-2}s^{-1}}\left(\frac{B_{mp}} {0.01{\rm
G}}\right)^2\left(\frac{v_{rel}}{138 \,\rm{km\;s^{-1}}}\right).
\label{eq:F_b}
\end{equation}
A fraction of this energy is converted to the kinetic energy of
the reconnection jets. The speed of the jets is on the order of
the local Alfv\'en speed $B_{mp}/\sqrt{4\pi\rho}\simeq 900$km
s$^{-1}\left(\frac{B}{0.01{\rm G}}\right)
\left(\frac{\rho}{10^{-21}{\rm g cm^{-3}}}\right)^{-1/2}$, where
$\rho=10^{-21}$ g/cm$^3$ is the typical mass density at $r\approx
10R_{\odot}$ (e.g. Suzuki \& Inutsuka 2006). We assume that the
same energy per unit mass is converted to the kinetic energy of
the electrons. Therefore the electrons are moving faster than the
ions by a factor of the square root of the ion-electron mass ratio
($\approx 40$). As a result, $v_b \approx 3\times 10^4$ km
s$^{-1}$ is a typical speed of the electron beam.

We denote the momentum flux and the energy flux of the electron
beam as $P_b$ and $F_b$, and their initial values as $P_{bi}$ and
$F_{bi}$, respectively. If we neglect thermal fluctuations of the
beam particles, the initial momentum flux and energy flux of the
electron beam are
\begin{eqnarray}
P_{bi}&=&\rho_{bi} v_b^2, \\
F_{bi} &=&\frac{1}{2}\rho_{bi} v_b^3,
\end{eqnarray}
where $\rho_{bi}$ is the initial beam mass density. In the present
numerical experiment, we restrict ourselves to the cases in which
$F_{bi} = 10^2$ and $10^4$ erg cm$^{-2}$s$^{-1}$. We adopt a
constant
$v_{b}=3\times 10^4$km s$^{-1}$, and vary $\rho_{bi}$ for
different $F_{bi}$. Note that according to eq.~(\ref{eq:F_b}) and
the parameters of the HD 179949 system, $F_{bi} = 10^2$erg
cm$^{-2}$s$^{-1}$ corresponds to $B_{mp} \approx 0.005-0.01$ G and
hence $B_* \approx 1$ G, and the larger energy flux $F_{bi} =
10^4$erg cm$^{-2}$s$^{-1}$ corresponds to $B_{mp} \approx
0.05-0.1$ G and thus the stronger stellar field $B_* \approx 5$ G.

We assume that the electron beam initially streams freely along
the flux tube at a constant $v_b$. As they approach the parent
star, the beam particles start to interact with the dense
background ions through Coulomb collisions and become more
concentrated due to the flux-tube convergence (see the next
subsection). The collisions lead to the heating of surrounding
plasma. The heating becomes efficient when the mean free path of
the beam is comparable to the local density scale height $H_\rho$.
The mean free path of an electron colliding with a pool of thermal
ions is given by $l_{\rm mfp} \approx 3\times 10^4 {\rm km}
\left(\frac{v_b}{3\times 10^4 {\rm
km\;s^{-1}}}\right)^4\left(\frac{n}{10^9 {\rm
cm^{-3}}}\right)^{-1}$ \citep{Braginskii}, where
$n=10^{9}$cm$^{-3}$ is the typical density at the upper transition
region or the lower corona (e.g. see SI05, or refer to the density
profile in Figure 1 to be discussed in \S3). Roughly speaking, for
$v_b=3\times 10^4$ km/s, $l_{\rm mfp} \lesssim H_\rho$ when
$n>10^9$cm$^{-3}$. Therefore, we anticipate the inward-propagating
electron beam to heat up ambient media from the lower corona to
the upper chromosphere. This allows us to assume that the incoming
electron beam starts dissipating from $r_{\rm max} =1.1R_{\odot}$
(lower corona) to $r_{\rm min}=1.001R_{\odot}$ (upper
chromosphere).

We model the energy flux of the beam to decrease inwardly
according to
\begin{equation}
F_{\rm b} \propto \left(\frac{r-r_{\rm min}}{r_{\rm max}-r_{\rm
min}} \right)^k , \label{eq:F_b_diss}
\end{equation}
where $k$ is a parameter that describes the spacial distribution
of the heating. $k=1$ corresponds to constant volumetric heating;
namely, the heating rate per unit mass is higher in the upper
region (i.e. corona) than in the lower region (i.e. chromosphere).
If $k= 0.1$, the heating rate per unit mass is more uniformly
distributed. In this work, we adopt $k=0.1$ to resemble the
situation of a constant beam-heating rate per unit mass. We also
assume that the momentum flux of the beam $P_b$ dissipates in the
same manner as that described by eq.~(\ref{eq:F_b_diss}) for the
energy flux. This implies that although the beam velocity $v_b$
does not decay in our dissipation model, the beam density $\rho_b$
declines, meaning that more and more beam electrons have been
transformed into thermal electrons as the beam progresses
downwards in the dissipation region.

The beam heating at the footpoint of one open field line occurs
when the planet's magnetosphere is crossing the field line. Hence,
by means of eq.~(\ref{eq:R_mp}), the beam heating proceeds on the
timescale
\begin{equation}
t_{beam}={2 R_{mp} \over v} \approx 80\, {\rm mins} \left( {R_p
\over R_J} \right) \left({v\over 138\,{\rm km/s}} \right)^{-1}
\left[ \left({B_p \over 1\,{\rm G}} \right) \left( {B_* \over
1\,{\rm G}} \right)^{-1} \left( {a \over 10R_{\odot}} \right)^2
\left( {R_* \over R_{\odot}} \right)^{-2} \right]^{1/3}.
\label{eq:t_heat}
\end{equation}

\subsection{Stellar Atmosphere Model}
The stellar atmosphere model is based on the 1-D
magnetohydrodynamic simulation with radiative cooling and thermal
conduction in an open flux tube (SI05). In the original
simulation, the heating is given by the nonlinear dissipation of
the Alfv\'en waves excited by the granulations at the photosphere.
In the case of $B_*=1$ G, we set a rms average amplitude
$<dv_{\perp}> \sim 1.8$ km/s at the photosphere, which is
estimated from the scaling with the surface convective flux
(Suzuki 2007) from the Sun (SI05). In the case of the stronger
field $B_*=5$ G, the larger rms velocity fluctuation $<dv_{\perp}>
\sim 3.6$ km/s is used for the experiment, which is twice the
value for the $B_*=1$ G case.
We neglect
the effect of rotation for simplicity in order to focus mainly on
the energetics influenced by the electron beam.


When an inward-propagating electron beam described in the above
subsection is taken into account, the momentum and energy
equations in a 1-D open flux tube are modified to (cf. SI05)
\begin{equation}
\label{eq:mom} \rho \frac{d v_r}{dt} = -\frac{\partial p}{\partial
r} - \frac{1}{8\pi r^2 f}\frac{\partial}{\partial r}  (r^2 f
B_{\perp}^2) + \frac{\rho v_{\perp}^2}{2r^2 f}\frac{\partial
}{\partial r} (r^2 f) -\rho \frac{G M_*}{r^2} - \frac{\partial
P_{\rm b}}{\partial r} ,
\end{equation}
and
\begin{eqnarray}
\label{eq:eng} \rho \frac{d}{dt}(e + \frac{v^2}{2} +
\frac{B^2}{8\pi\rho} - \frac{GM_*}{r}) + \frac{1}{r^2 f}
\frac{\partial}{\partial r}[r^2 f \{ (p + \frac{B^2}{8\pi}) v_r -
\frac{B_r}{4\pi} (\mbf{B \cdot v})\}] \nonumber \\
+ \frac{1}{r^2
f}\frac{\partial}{\partial r}(r^2 f F_{\rm c})
+ \frac{1}{r^2 f}\frac{\partial}{\partial r}(r^2 f F_{\rm b}) +
q_{\rm R} = 0,
\end{eqnarray}
where $\rho$, $\mbf v$, $p$, $e$, and $\mbf B$ are the density, velocity, pressure,
specific internal energy, and magnetic field strength, respectively; the subscripts
$r$ and $\perp$ denote radial and tangential components, respectively; $d/dt$ and
$\partial/\partial t$ denote Lagrangian and Eulerian derivatives, respectively; $G$
and $M_*$ are the gravitational constant and the stellar mass, respectively; $F_c$
is thermal conductive flux; $q_R$ is the radiative cooling and $f$ is a
super-radial expansion factor. We assume that the super-radial expansion (in
addition to the radial expansion $\propto r^2$) is a factor of 240 and 480 from the
photosphere to $\approx 2 R_{\odot}$ in the cases of $B_*=1$ and 5 G, respectively.
It then follows that to give $B_*=$ 1G and 5G, the radial magnetic field strength
at the footpoint of the flux tube at the photosphere $B_{ph}$ is 240G and 2400G,
respectively.

Note that on account of the convergence of the flux tube toward the central star,
the energy flux of the incoming electron beam increases due to the areal focusing.
In the case of $B_*=1$G, the flux tube converges by a factor of $\approx$14,600 (a
factor of $\approx 61$ from the radial convergence and another factor of 240 from
the super-radial convergence) from $7.8R_*$ (the planet's orbit) to $1R_*$. On the
other hand, the converging factor of the flux tube in the $B_*=5G$ case is twice as
large. In the absence of dissipation, the energy flux of the beam increases by the
same factor as the converging factor of the flux tube. In contrast, the same
enhancement effect of the electron beam pressure on stellar thermal ions is almost
negligible in our calculations because of the small mass of an electron.

\section{Results of the Numerical Experiment}


We conduct the numerical experiment to study the effects on the
stellar atmosphere due to the electron beam with the initial
energy fluxes for the two $B_*$ cases as described in the
preceding session. The experiment is first carried out in the
absence of the beam until the simulated atmosphere attains a
quasi-steady state. The beam heating is then added afterwards.

Figure 1 shows the snapshot structures of the stellar atmosphere
for $B_*=$ 1 and 5 G denoted respectively by ``small" and ``large
$B$ and $dv$", and compares the results with and without the
electron-beam injection.
In the $B_*=1$G case, the results for $F_{bi}=10^2$ erg cm$^{-2}$
s$^{-1}$ (blue curves) are almost identical to those for the
no-beam case (green curves). In contrast, the atmospheric
structures change noticeably in the larger $B_*$ ($=5$G) and
$<dv_\perp>$ ($=3.6$ km/s) cases, as illustrated by the red and
black curves.
In the absence of the beam heating, the stronger dissipation
arising from the larger $<dv_\perp>$ raises the chromospheric
temperature and therefore causes the chromosphere to evaporate and
expand (see the red curve in the middle panel). The local density
scale height increases and the density drops more slowly with $r$.
Therefore, the locations of the transition region lie at higher
altitudes in the larger $<dv_\perp>$ case. A hot and dense region
forms in the chromosphere, which we term as ``a warm region" (i.e.
$T>10^4$K, see the bottom panel). It then becomes difficult to
further heat up the warm region to the coronal temperature (i.e.
$\gtrsim 10^6$ K) as a result of the efficient radiative cooling
at $T\lesssim 10^5$ K \citep{LM90}. However, when the beam energy
of $F_{bi}=10^4$ erg cm$^{-2}$ s$^{-1}$ is added, the chromosphere
is further heated up and thus an even larger warm region can form
at the location of $r=1.001 - 1.01 R_*$ (see the black curve in
the bottom panel).
The black curve in the middle panel indicates that the density is
on the average much higher in the beam-heated region.
Consequently, the heating rate per unit mass becomes smaller and
the temperature of the coronal region at $r \gtrsim 1.01R_*$ then
becomes lower in the case of $F_{bi}=10^4$ erg cm$^{-2}$ s$^{-1}$
than in the case of $F_{bi}=0$ (see the bottom panel). The sharp
transition region disappears.

The top panel of Figure 1 shows the radial profile of the stellar
wind velocity $v_r$. The winds in the outer region ($r\gtrsim
2R_*$) are faster in the large $B_*$ case. $B_*$ equals
$B_{ph}/f$, which in fact specifies the flux tube properties. The
larger the factor $B_{ph}/f$ is, the more the wave energy
dissipates in the outer region of the atmosphere, leading to
faster winds \citep{Kojima,Suzuki06}. However, the winds on the
average are not significantly amplified by the beam heating, as
indicated by the large overlap between the red and black curves in
the plot. The wind speed at the planet's orbit is $\approx 300$ km
s$^{-1}$, which is lower than the Alfv\'{e}n speeds $\sim 500$ and
1000 km s$^{-1}$ there in our cases for $B_*=1$ and 5 G. That is,
the planet lies inside the Alfv\'{e}n radius.

Having described the snapshot structures, we should note that the
thermal properties of the stellar atmosphere actually fluctuate
with time owing to wave propagation and dissipation. As a result,
the location and thermal properties of the warm region fluctuate
with time as well. Figure 2 shows the evolution of the radiative
fluxes arising from different temperature components in the beam
and no-beam cases. In general, all of the radiative fluxes
fluctuate with time. Nevertheless for the case of $B_*=1$G and
$<v_\perp>=1.8$ km/s, the electron beam of $F_{bi}=10^2$ erg
cm$^{-2}$ s$^{-1}$ gives rise to only tiny effects on the
radiations compared to the no-beam case, as expected from the
snapshot structures shown in Figure 1. In contrast, the case of
$F_{bi}=10^4$ erg cm$^{-2}$ s$^{-1}$ corresponding to $B_*=5$G
exhibits noticeable differences from the no-beam case, which can
also be expected from the snapshot structures in Figure 1. The
radiative fluxes from the hot chromosphere (7000-20000K) and the
transition region (20000-$5\times 10^5$K) are greatly enhanced by
the warm region in the vicinity of $r=1.001-1.01R_*$. Since the
warm region is on the average denser than the usual chromosphere
and transition region in $r\gtrsim 1.003 R_*$ (see Figure 1), the
radiative flux corresponding to the chromospheric temperature is
on average increased by a factor of $\sim10$ and the radiative
flux associated with the transition-region temperature is
intensified by a factor of $\sim100$.

The bottom panel of Figure 2 shows that the beam of $F_{bi}=10^4$
erg cm$^{-2}$ s$^{-1}$ associated with $B_*=5$G is able to enhance
the radiation from the hot gas of $T>5\times 10^5$K occasionally
by a factor of $\gtrsim 100$. This radiation is mainly from the
warm region that intermittently develops around
$r=1.001-1.003R_*$.
With this beam heating, the temperature of the warm region can
sometimes go up and down between $10^5$K and $10^6$K. This
temperature variability is due to the thermal instability
\citep{LM90,Suzuki07}. The electron beam is energetic enough to
continuously heat up the warm region, while the wave dissipation
heats it up in a stochastic manner. A small change of the wave
heating rate triggers violent fluctuations of temperature in the
thermally unstable regime of $10^5 < T < 10^6$K.

As is shown in Figure \ref{fig2}, the thermal response of the
stellar atmosphere to the onset of the beam-heating to evolve to
the hot state is nearly instantaneously. In \S \ref{sec:ebi}, we
estimated the duration of the electron beam as $\sim 100$ min,
which is rather short in comparison with the simulation time
presented in Figure \ref{fig2}. To estimate the area of the hot
spot, it is equally important to examine how long the hot
atmosphere can be sustained after the electron beam passes by.
Figures \ref{fig4} \& \ref{fig5} show the results for the $B_*=5$G
case
but the beam is switched off at $t=109$ min. Figure \ref{fig4},
which presents the snapshot structure at $219$ min, namely 110 min
after the beam stops, illustrates that while the warm region in
the chromosphere is still present as shown by the temperature
profile, its density has
dropped to the original lower-density level. Figure \ref{fig5}
shows that the radiative fluxes from the hot chromosphere (top)
and transition regions (middle) have declined for $\approx$ 50-60
minutes since the beam is switched off, which indicates that the
atmosphere takes only a fraction of the heating time to revert to
the normal state in the absence of the electron beam. Therefore,
the size of the hot spot estimated in terms of the projected area
of the planet's magnetosphere along a flux tube onto the
chromosphere is a reasonable approximation.

\citet{Shk05,Shk08} found that in the case of HD 179949, the planet induced Ca II
flux averaged over the stellar disk is $\sim 1.5\times 10^5$ ergs cm$^{-2}$
s$^{-1}$, which amounts to an increase of a factor of about 1.04 compared to the
non-planet induced Ca II emissions. Hence the Ca II emissions with no planet
induced component is $\sim 1.5/0.04 \times 10^5 = 3.7 \times 10^6$ erg cm$^{-2}$
s$^{-1}$, which is on the similar order of the value given from the model shown by
the red-dotted line in Figure \ref{fig2}. In this sense, the model for $B_*=5 $G
and $<dv_\perp>=3.6$ km/s may mimic a chromospheric condition similar to that of HD
179949, although our cooling function $q_R$ for the chromospheric radiation is
based on the observation of the Sun \citep{AA89}. Having obtained the radiative
flux $\sim 10^{6-7}$ erg cm$^{-2}$ s$^{-1}$ from the hot chromosphere in the case
for $B_*=5$G and $<dv_\perp>=3.6$ km/s, we can estimate the luminosity of the hot
spot. In our model, the cross-sectional area of the converging flux tube in the
chromosphere is about 8000 times smaller than the planet's magnetosphere. In other
words, the area of the hot spot in the chromosphere is given by
\begin{equation}
A_{chromo}\approx 10^{17-18} \left( {R_{mp} \over 5 R_J} \right)^2
\,{\rm cm^2},
\end{equation}
which when multiplied by the chromospheric flux gives the luminosity of the
chromospheric hot spot $\approx 10^{23-25}$ erg/s. This total chromospheric
emission is still 2-4 orders of magnitude weaker than the observational Ca II
emissions for HD 179949.






\section{Summary and Discussions}

By conducting a numerical experiment, we study the thermal
response of the atmosphere of a solar-type star to the dissipation
of an injected electron beam at the coronal base. The experiment
is carried out based on the framework of the 1-D
magnetohydrodynamic simulation by SI05 with non-linear wave
dissipation, radiative cooling, and thermal conduction. We assume
that the magnetic stress due to the orbital motion of the planet
relative to the stellar coronal fields generates an electron beam,
which in turn funnels along the stellar open field lines to the
central star. As the beam travels inwards, the energy flux of the
incoming electron beam is intensified by the areal focusing of the
super-radially converging open flux tube.

We use the stellar parameters of HD 179949 as an illustrative
example but ignore possible magnetic properties arising from its
stellar rotation. When the average stellar field at the
photosphere $B_*$ is about 1 G and the average amplitude of the
wave velocity $<v_\perp>$ is about 1.8 km/s, the stellar
atmosphere is not considerably altered after the beam dissipation
is turned on. In contrast, when $B_*=$5 G and $<v_\perp>=3.6$
km/s, we find that a warm region forms in the chromosphere. The
warm region becomes substantially hotter and denser once the
electron-beam heating is switched on. As a result, the
beam-intensified warm region enhances the chromospheric radiative
flux by a factor of $\sim10$, and the radiative flux corresponding
to the temperature of the transition region by a factor of
$\sim100$. The warm region can also intermittently contribute to
the radiative flux associated with the coronal temperature by a
factor of $\sim100$ due to the thermal instability \citep{LM90}.
In other words, the planet-induced radiations are not
perturbations in the local region of the hot spot compared to the
normal state of the stellar emissions. However, owing to the small
area of the heating spot, the total luminosity of the beam-induced
chromospheric radiation is 2-4 orders of magnitude smaller than
the observed Ca II emissions from HD 179949.

The energetics of the planet-induced emissions becomes a more
serious problem in our numerical experiments when explaining the
statistical results of X-rays from planet-host stars: stars with
close-in giant planets are on average more X-ray active by a
factor $\sim 4$ than those with planets that are more distant
\citep{Kashyap08}. Since the typical X-ray luminosity from a
solar-type stars is $\gtrsim 10^{27-28}$ erg/s, the planet-induced
X-ray inferred from the statistical analysis is actually even
stronger than the planet-induced Ca II emissions from HD 179949.
Our simulation results show that an $\sim 100$ times enhancement
in X-ray due occasionally to the thermal instability of the small
warm region contributes only an even more negligible perturbation
to the total X-ray emission, rather than being comparable to it.

We note that our open-field model for the chromospheric emissions
is different from those occurring on the Sun where the emissions
come primarily from the solar plage regions in closed magnetic
loops. Needless to say, our 1-D numerical experiment restricts
ourselves to exclude any mechanical and thermal influence of the
heating area on neighboring open fields and closed magnetic loops.
As such, our results leave an open question as to whether the
thermal instability or any other magnetic instabilities (e.g.
Lanza 2008; Ishikawa et al. 2008) can be further triggered around
the beam-heated spot to liberate more energy.

In our numerical experiment $B_*$ has been taken to be 1 and 5 G,
which is consistent with the field measurements via
spectrapolarimetry for the similar spectral type dwarf $\tau$
Bootis \citep{Catala}. However, the field strength inferred from
the Stokes V observation may be underestimated due to the
cancellation of circular polarization arising from the opposite
directions of $B_*$ along the line of sight. It is normally
expected that a faster rotator and therefore a stronger X-ray
emitter may possess stronger $B_*$ than the Sun
\citep{Ruedi,Gudel}. For instance, $B_*$ of HD 179949 has been
assumed to be $\approx$ 8-9$\times$ solar value by scaling with
the X-ray flux \citep{Saar}. Furthermore, \cite{MJ} related $B_*$
and the filling factor $f_s$ to the Rossby number $Ro$ ($\equiv
P_{rot}/\tau_c$, where $P_{rot}$ is the rotation period and
$\tau_c$ is the convection turnover time) of a dwarf star. In the
case of HD 179949, $P_{rot}$ has been suggested to be about 7 days
\citep{WH04,Shk08} and $\tau_c \approx 4.287$ days may be inferred
from its color B$-$V=0.503 (Noyes et al. 1984). Thus following
\cite{MJ}, we obtain the average field $Bf_s=37.6$ G and the
filling factor $f_s=0.02$. If we assume that these magnetic
properties are contributed mainly from the open-field region of
our model, then $Bf_s = 37.6$ G is equivalent to $B_*$ and the
filling factor $f_s=0.02$ may correspond to the super-radial
expansion factor $1/f$ in our experiment. In reality, some of the
contribution may come from closed-field regions \citep{MJ}, which
introduces additional complexity.

The other free parameter that governs our numerical results is the
power spectrum of MHD waves. The stellar macroturbulent velocity
$v_{mac}$ for HD 179949 is expected to be 2 times larger than that
for the Sun \citep{SO97,Saar}. In our experiment, we adopt a
larger $<dv_{\perp}>$ for the stronger $B_*$ case.
However, the correlation between $v_{mac}$ and $<dv_{\perp}>$ is
not tested in this work and how these velocities are associated
with $B_*$ is not modelled in our numerical experiment. After all,
in view of all of the uncertainties and complexities mentioned
above, our results serve only as the fiducial examples for future
studies. The numerical experiment covering a broader parameter
space coupled with more magnetic-field measurements will be
essential to further diagnose the problem.

The effect of the centrifugal force is not implemented in the
present numerical experiment. While it is a reasonable
approximation for thermally driven winds from a slow rotator like
the Sun, the magneto-centrifugal winds play an equally important
role in accelerating stellar winds for a star rotating $\sim 10$
times faster than the Sun \citep{BM76,WS93,Pre06}.
The centrifugal force of a 7 day-period star is $\sim 1/5$ that of
a 3 day-period star, as the force is proportional to the square of
the rotation frequency. The effect of the centrifugal force on the
stellar atmosphere will be investigated in a future work.

While most of the previous magnetohydrodynamic simulations and studies have focused
on the planet's side for the magnetic interactions (Ip et al. 2004; Preusse et al.
2006, 2007; cf. Laine et al. 2008 for a young hot Jupiter), our numerical
experiment makes an attempt to investigate
how a stellar atmosphere down to
the photosphere in the open field region responds to the dissipation likely from a
hot Jupiter. The current model is simplified in such a way that we prescribe the
energy dissipation and ignore stellar rotation. Nevertheless, the dissipation
contributed by the planet is described only by the energy flux $F_b$ at the coronal
base, meaning that $F_b$ is not necessarily specific only to an incoming electron
beam but can be in the form of other dissipative sources (e.g. damping of Alf\'ven
waves) with proper modification. Furthermore, our simulation lays the framework to
extend the calculations for other types of dwarf stars if the corresponding
magnetic wave amplitudes and power spectra are specified.




\acknowledgments

We wish to dedicate this work in memory of Chi Yuan who passed
away on July 24, 2008. Without him, the authors would not have met
each other and finally made the work possible. We are grateful to
N. Phan-Bao and E. Shkolnik for useful discussions. We also thank
the anonymous referee for valuable comments to improve the paper.
P.-G. Gu is supported by the NSC grants in Taiwan through NSC
95-2112-M-001-073MY2 and 97-2112-M-001-017. T. K. Suzuki is
supported in part by Grants-in-Aid for Scientific Research from
the MEXT of Japan, 19015004 and 20740100 and Inamori Foundation.

\clearpage



\begin{figure}
\epsscale{.60}
\plotone{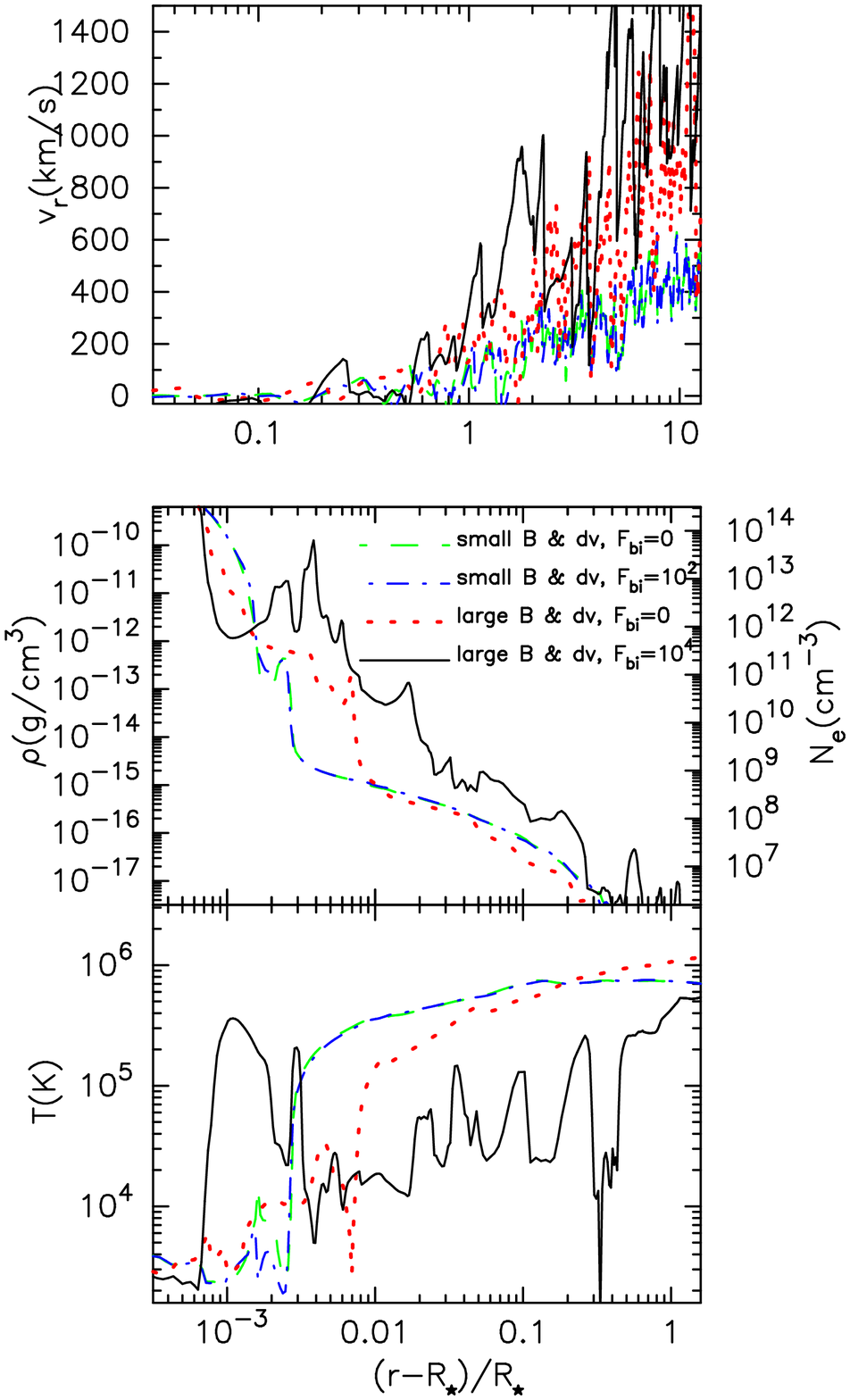} \caption{Snapshot structures of the radial
velocity, density, and temperature for the different beam energy
fluxes at 142 minutes after the electron beam is switched on. The
green dashed and blue dash-dotted curves are the results of
$F_{bi}=0$ and $10^2$erg cm $^{-2}$s$^{-1}$ for $B_*=1$ G, whereas
the red dotted and black solid curves are the results of
$F_{bi}=0$ and $10^4$ erg cm$^{-2}$s$^{-1}$ for $B_*=5$ G. Note
that the vertical scale of the radial velociy is different from
that of the density and temperature. \label{fig1}}
\end{figure}

\clearpage


\begin{figure}
\epsscale{.70}
\plotone{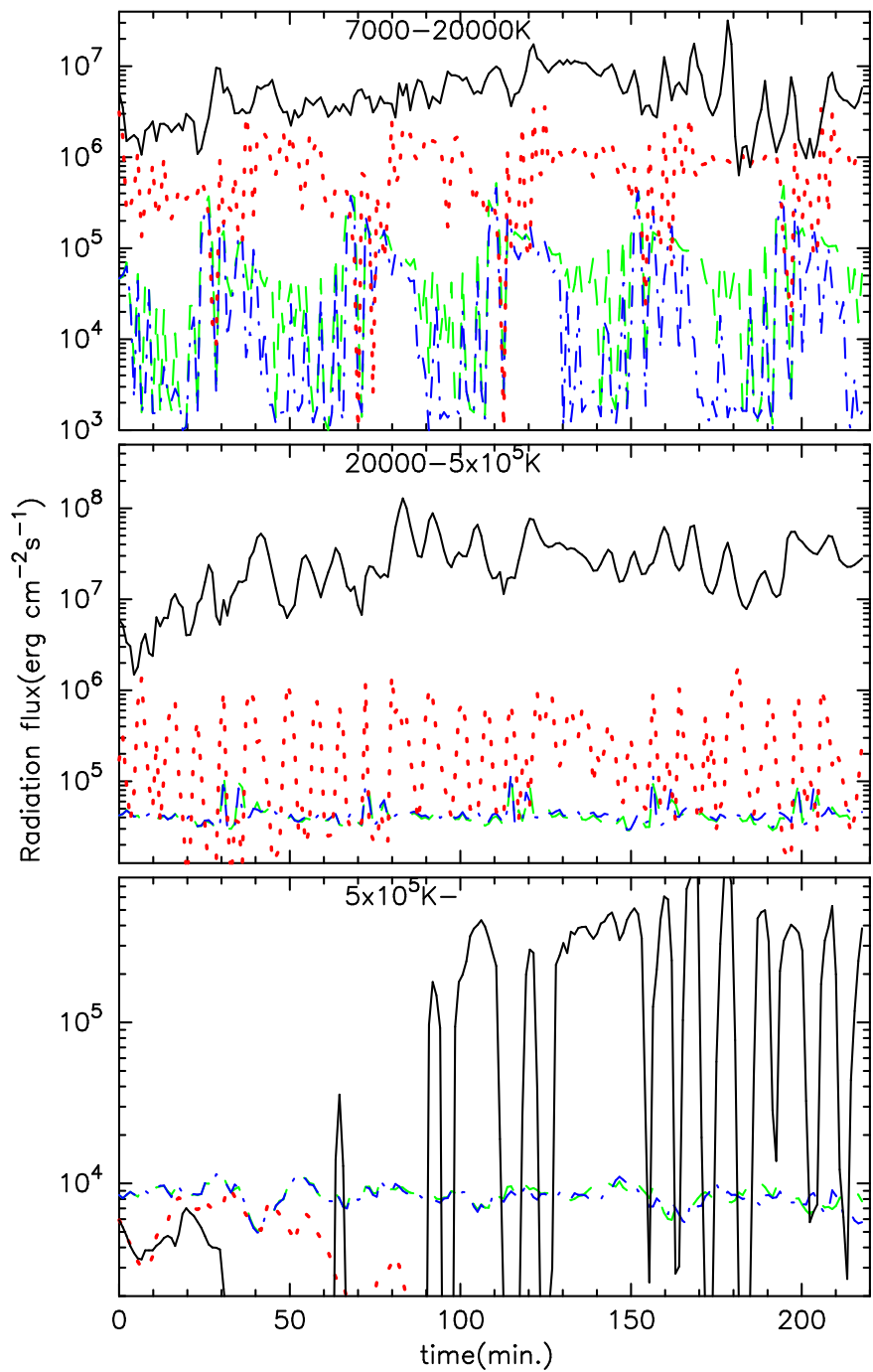} \caption{Comparison of the radiative flux arising
from different temperature components. The top, middle, and bottom
panels show the radiative flux from the gas at the temperatures
corresponding to the hot chromosphere, transition region, and
corona. The cases for $B_*=1$ and 5 G in the absense and presence
of the beam heating are plotted with the same color coding as
those shown in Figure \ref{fig1}. \label{fig2}}
\end{figure}


\begin{figure}
\epsscale{.80}
\plotone{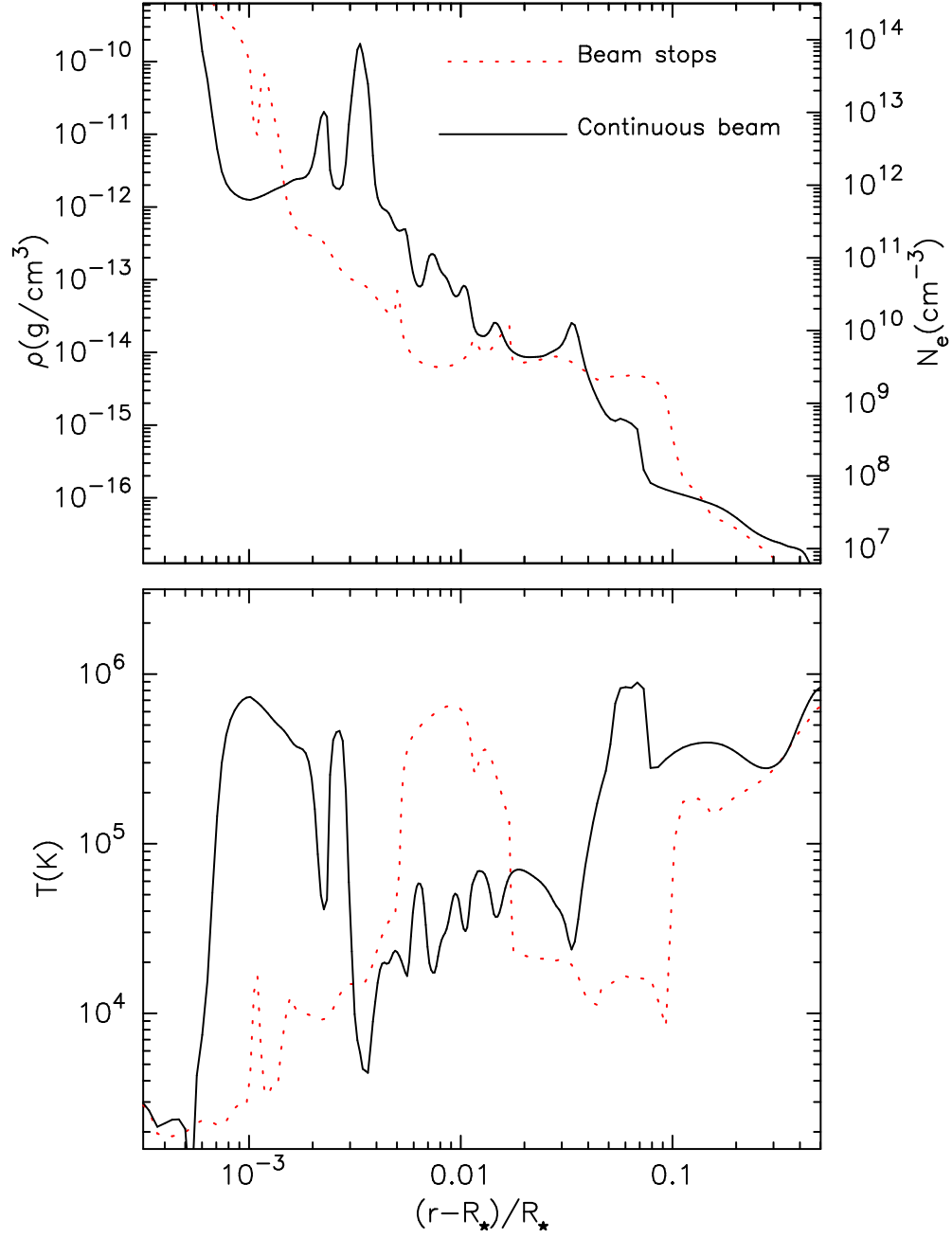} \caption{Comparison of snapshot structures of the
density and the temperature at $t=219$ minutes for the case with
the continuous beam of the energy flux $F_{\rm bi}= 10^4$erg
cm$^{-2}$ s$^{-1}$ (black solid curve) and the case with the beam
switched off at 109 minutes (red dotted curve). \label{fig4}}
\end{figure}

\begin{figure}
\epsscale{.70}
\plotone{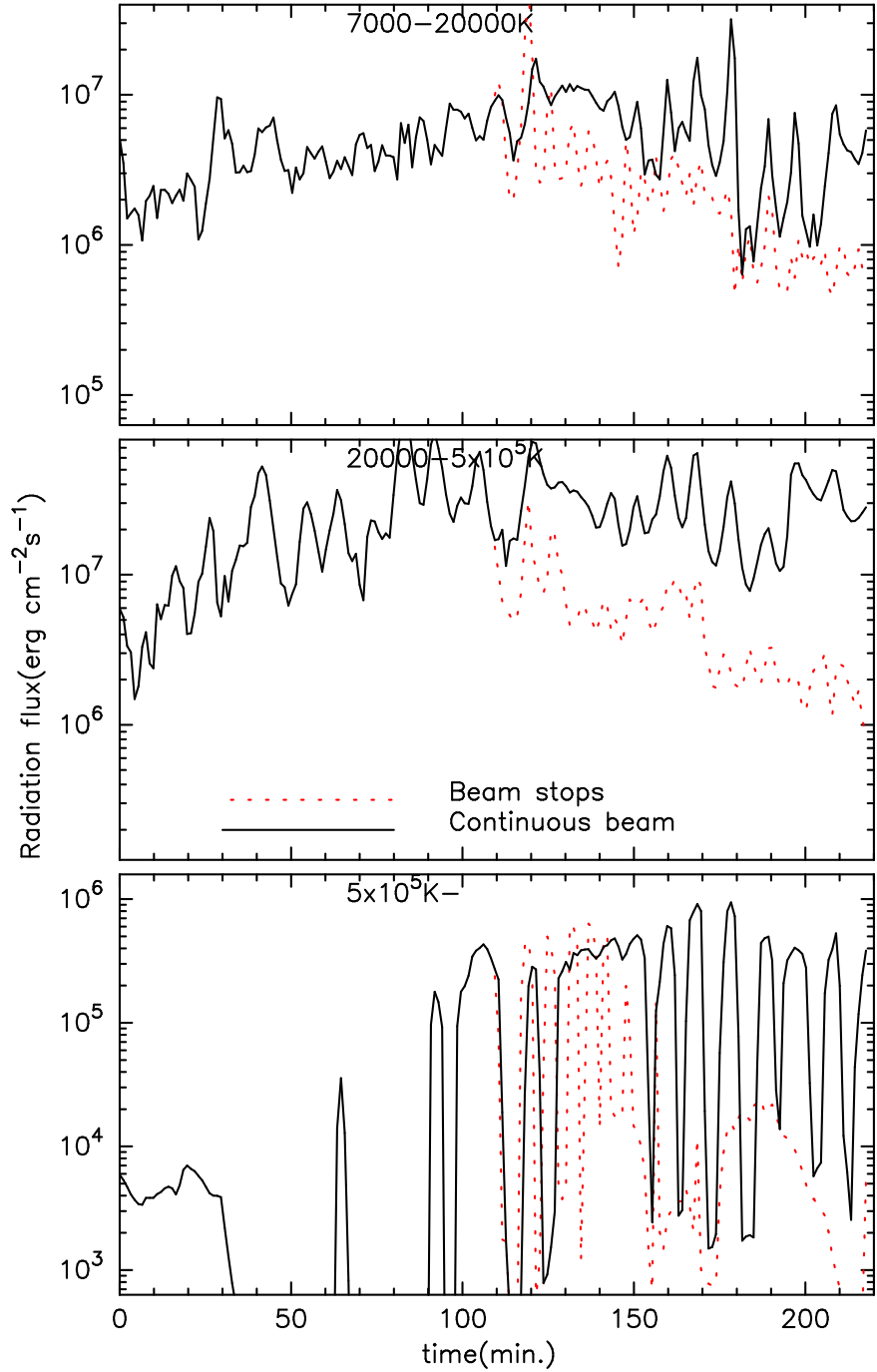} \caption{Comparison of the radiative flux for the
case with the continuous beam (solid) and the case with the beam
switched off at 109 minutes (red dotted). The three panels show
the radiation from the gas at the temperatures corresponding to
hot chromosphere (top), transition region (middle), and corona
(bottom). \label{fig5}}
\end{figure}









\end{document}